\documentclass[floatfix,groupedaddress,10pt,twocolumn]{revtex4-1}
\usepackage{graphicx}
\usepackage{graphics, color}
\usepackage{rotating}
\usepackage{setspace} 
\usepackage{lipsum}
\usepackage{overpic}
\usepackage{float}
\usepackage[english]{babel}
\usepackage{hyperref}
\usepackage[usenames,dvipsnames]{xcolor}
\usepackage{adjustbox}
\usepackage{amsmath,amsfonts,amssymb,verbatim,float,mathtools}
\usepackage{physics}
\usepackage[titletoc,title]{appendix} 
\graphicspath{{pics/}}
\definecolor{Gray}{gray}{0.4}
\hypersetup{%
	pdfborder = {0 0 0}}

\begin{document}		
	\title{Phases of driven two-level systems with nonlocal dissipation}
	
	\author{C. D. Parmee}
	\affiliation{T.C.M. Group, Cavendish Laboratory, University of Cambridge, JJ Thomson Avenue, Cambridge, CB3 0HE, U.K.}
	\author{N. R. Cooper}
	\affiliation{T.C.M. Group, Cavendish Laboratory, University of Cambridge, JJ Thomson Avenue, Cambridge, CB3 0HE, U.K.}
	\begin{abstract}
		We study an array of two-level systems arranged on a lattice and illuminated by an external plane wave which drives a dipolar transition between the two energy levels. In this set up, the two-level systems are coupled by dipolar interactions and subject to nonlocal dissipation, so behave as an open many-body quantum system. We investigate the long-time dynamics of the system at the mean-field level, and use this to determine a phase diagram as a function of external drive and detuning. We find a multitude of phases including antiferromagnetism, spin density waves, oscillations and phase bistabilities. We investigate these phases in more detail and explain how nonlocal dissipation plays a role in the long-time dynamics. Furthermore, we discuss what features would survive in the full quantum description.
	\end{abstract}
	\date{\today} 
	\maketitle

	\section{Introduction}
	A recurring problem in physics concerns the interaction of an electromagnetic wave with a medium formed from an array of polarisable particles. If these particles are two-level quantum systems driven close to resonance, then collective effects can arise due to the strong resonant dipole-dipole interactions provided the average interparticle spacing is smaller than the dipolar transition wavelength. These collective effects give significant deviations in the behaviour of the medium compared to one formed of non-interacting scatterers. Key differences include the emergence of Lamb shifts, where interactions modify the two-level transition energies \cite{Keaveney2012,Friedberg1973}, and the formation of super- and sub-radiant modes, where the dipole-dipole interactions enhance or suppress decay of excitations \cite{Radiation1954,Gross1982}. Understanding how these collective effects alter the response of a medium is still an ongoing topic of research.  
	
	A natural place to study strong dipole-dipole interactions is in cold atom systems, where a high level of control of the interaction strength and atom spacing is possible. Much work has already been carried out on theoretical understanding of light scattering through cold atom gases \cite{Zhu2016,Fofanov2011,Sutherland2016,Bienaime2012,Bienaime2013,Lee2016} with some of these effects being realised experimentally \cite{Guerin2016,Jenkins2016}. Most work focuses on the low light intensity limit, where interactions between excitations is negligible and the full quantum model simplifies to a problem of classical scatterers. In these models, the collective effects can be exploited, especially when the atoms are arranged periodically on a lattice, leading to effects such as electromagnetically induced transparency interferences \cite{Bettles2015a,Feng2017}, long-time excitation storage in subradiant modes \cite{Facchinetti2016} and enhanced optical cross sections \cite{Bettles2016a}. 
	
	However, much less work has been done beyond low intensities, and has been largely limited to small system sizes \cite{Levi,Olmos2014,Yu2014,Yu2015}. In the case where two atomic transitions are isolated, the problem of light scattering from a cold atomic gas can be mapped to a driven-dissipative spin$-1/2$ system. At moderate to high intensity drive, these spin systems show novel phases such as optical bistability, Anti-ferromagnetic (AFM) and Spin Density Wave (SDW) order and even oscillations (OSC) where the spins oscillate well into the long-time limit \cite{Chan2015,Lee2011,Lee2013}. Such systems are also realisable in coupled cavities \cite{Wilson2016,Armen2006,Jin2014}.
	
	Here, we study the properties of a driven cold atomic ensemble beyond the regime of low intensity by employing a numerical mean-field  analysis of a large number of two-level systems on a lattice. We establish the open system phase diagram in a 1D geometry and find examples of all the phases mentioned above. We also explain how these phases arise due to interactions and the presence of nonlocal dissipation which causes super/subradiant decay. Previous work has given evidence of bistabilities for uniform mean-field states when examining small systems \cite{Olmos2014}. In this paper we investigate larger systems with a different dipole orientation. Our work shows the emergence of spatial and temporal phases that were not evident in other studies.	
	
	The paper is organised as follows. In section \ref{sec:setup} we set up our model. In section \ref{sec:phasediagram}, we establish the mean field phase diagram with quantum checks in \ref{sec:beyondmeanfield}. Finally, in \ref{sec:discussion} we discuss our results and possible experimental realisations before drawing conclusions in \ref{sec:conclusions}.
	
	\section{Model}
	\label{sec:setup}
	We consider a large number, $N$, of two-level systems fixed in position in a deep 1D optical lattice to form a 1D array. The two-level systems are illuminated with a uniform plane wave and coupled to one another by resonant dipole-dipole interactions. The system is also coupled to the electromagnetic field in free space, which acts as a Markovian environment and allows the dipoles to decay. The resultant Master equation is given by \cite{James1993,Bettles2016b}
	\begin{equation}\label{MasterEq}
	\begin{split}	\dot{\hat{\rho}}_{N}(t)=&-\frac{i}{\hbar}\left[\hat{H}_{\rm sys}+\sum_{i\neq l}^{N}\hbar V_{il}\hat{\sigma}_i^+\hat{\sigma}_l^-,\hat{\rho}_{N}(t)\right]\\
	&+\sum_{i,l}^{N}\frac{\Gamma_{il}}{2}\left[2\hat{\sigma}_i^-\hat{\rho}_{N}(t) \hat{\sigma}_l^+-\left\lbrace \hat{\sigma}_l^+\hat{\sigma}^-_i,\hat{\rho}_{N}(t)\right\rbrace\right],	\end{split}
	\end{equation}
	where the square brackets represent a commutator, curly brackets represent the anti-commutator and $\hat{\sigma}_l^{\pm}=\hat{\sigma}_l^{x}\pm i\hat{\sigma}_l^{y}$ where $\hat{\sigma}_l^{\alpha}$ are the Pauli matrices on site $l$ with $\alpha=x,y$ or $z$. The on-site Hamiltonian is given by $\hat{H}_{\rm sys}=\hbar\Omega/2\sum_{i}^{N}\hat{\sigma}^x_i-\hbar\Delta/2\sum_{i}^{N}\hat{\sigma}^z_i$, where $\Delta=\omega-\omega_0$ is the detuning from the  two level transition energy, $\omega_0$, and $\Omega=2\textbf{d}.\textbf{E}/\hbar$ is the Rabi coupling determined by the dipole moment vector \textbf{d} and the electric field vector $\textbf{E}$. We consider an experimental set up where the wavevector of the drive, $\textbf{k}$, is perpendicular to the lattice and the electric field parallel to the lattice so that $\textbf{E}=E_0\boldsymbol{\hat{x}}e^{-iky}$. The dipole interactions and decay terms are then given by
	
	\begin{equation}
	\begin{split}
	&V_{il}=-\frac{3\Gamma}{2}\left(\frac{\sin\kappa r_{il}}{(\kappa r_{il})^2}+\frac{\cos\kappa r_{il}}{(\kappa r_{il})^3}\right) \\
	&\Gamma_{il}=3\Gamma\left(-\frac{\cos\kappa r_{il}}{(\kappa r_{il})^2}+\frac{\sin\kappa r_{il}}{(\kappa r_{il})^3}\right), \\
	\end{split}
	\end{equation}
	
	where the single atom decay rate is given by $\Gamma=|\textbf{d}|^2 \kappa^3/3\pi\epsilon_0\hbar$ and $r_{il}=|\textbf{r}_i-\textbf{r}_l|$ where $\textbf{r}_i$ are the positions of atom $i$ on the lattice. Note that $\Gamma_{ii}=\Gamma$ so there is local as well as nonlocal dissipation in the system.
	
	\begin{figure}
		\hspace*{-0.3cm}
		\includegraphics[scale=0.4,clip,angle=0]{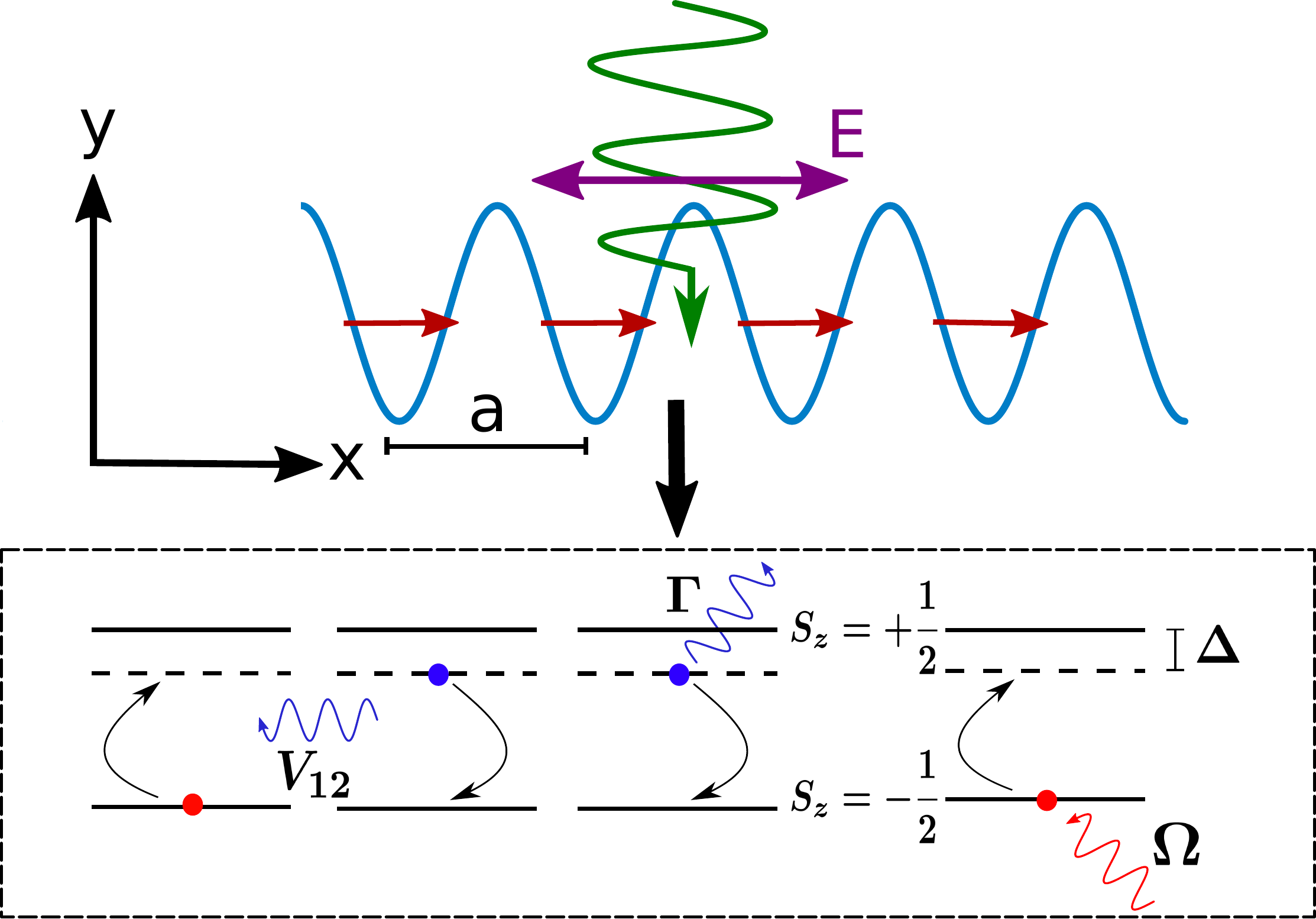}
		\caption{A schematic of  a 1D array of atoms under external drive. The electric field, shown by the purple arrow, is oriented parallel to the x axis and controls the orientation of the dipoles shown in red. The lattice spacing is denoted by $a$. The boxed image shows the microscopic picture of two-level systems interacting via photon exchange and dissipation, where the external drive controls the value of the Rabi coupling, $\Omega$.}
		\vspace{-0cm}
		\label{SystemModel}
	\end{figure}

	The parameter $\kappa a=2\pi a/\lambda$ is the ratio of the two-level transition wavelength, $\lambda$, to the lattice spacing $a$, and is important in determining the nature of the interactions and loss. If we consider $\kappa a \approx 0$, with $a$ fixed, then the system becomes closed, reducing to a quantum XY model with dipolar coupling and negligible dissipation. If we instead consider $\kappa a \approx 0$, with $\kappa$ fixed, then the interaction strength diverges and dissipation becomes all-to-all with $\Gamma_{il} = \Gamma$. In the opposite limit where $\kappa a\gtrsim2\pi$, the interactions become negligible and the dissipation becomes local (however, we will later find that $\kappa a\lesssim 1.2$ to observe interesting results). Throughout the rest of this paper, we work with $\kappa a=0.7$ which is well within these limits and allows us to see the effects of nonlocal interactions and dissipation. We also work in units where $\hbar=1$. 
	
	Analysing the behaviour for a large number of spins becomes intractable in the full quantum regime as the Hilbert space grows as $2^N$. To proceed, we make the Gutzwiller mean-field approximation, $\hat{\rho}_N\approx \otimes\hat{\rho}_i$ which results in ignoring quantum entanglement across lattice sites. Then, by taking the trace of Eq. \eqref{MasterEq} over all the sites except a given site $l$, we obtain the equations of motion as
	\begin{equation}\label{SpinEqs}
	\begin{split}
	\frac{dS_{l}^{x}}{dt}=&-\frac{\Gamma}{2} S^{x}_{l}-\Delta S^{y}_{l}-2\sum_{i(\neq l)}^{N}V_{il}S_{l}^{z}S_{i}^{y}+\sum_{i(\neq l)}^{N}\Gamma_{il}S_{l}^{z}S_{i}^{x}\\
	\frac{dS_{l}^{y}}{dt}=&-\frac{\Gamma}{2} S^{y}_{l}+\Delta S^{x}_{l}-\Omega S^{z}_{l}\\
	&+2\sum_{i(\neq l)}^{N}V_{il}S_{l}^{z}S_{i}^{x}+\sum_{i(\neq l)}^{N}\Gamma_{il}S_{l}^{z}S_{i}^{y}\\
	\frac{dS_l^{z}}{dt}=&-\Gamma (S_{l}^{z}+1/2)+\sum_{i(\neq l)}^{N}\Gamma_{il}(S_{i}^{x}S_{l}^{x}+S_{l}^{y}S_{i}^{y})\\
	&+\Omega S^{y}_{l}-2\sum_{i(\neq l)}^{N}V_{il}(S_{i}^{y}S_{l}^{x}-S_{l}^{y}S_{i}^{x}),
	\end{split}
	\end{equation}
	where $S^{\alpha}_l=\frac{1}{2}\Tr(\hat{\sigma}^{\alpha}_l\hat{\rho}_N)$ are the spin expectation values. We have solved the dynamics of the non-linear Eqs. \eqref{SpinEqs} and found the steady state solutions in the long-time limit.

	\section{Mean-Field Phase Diagram}
	\label{sec:phasediagram}
\begin{figure}
		\hspace*{-0.5cm}
		\includegraphics[scale=0.65,clip,angle=0]{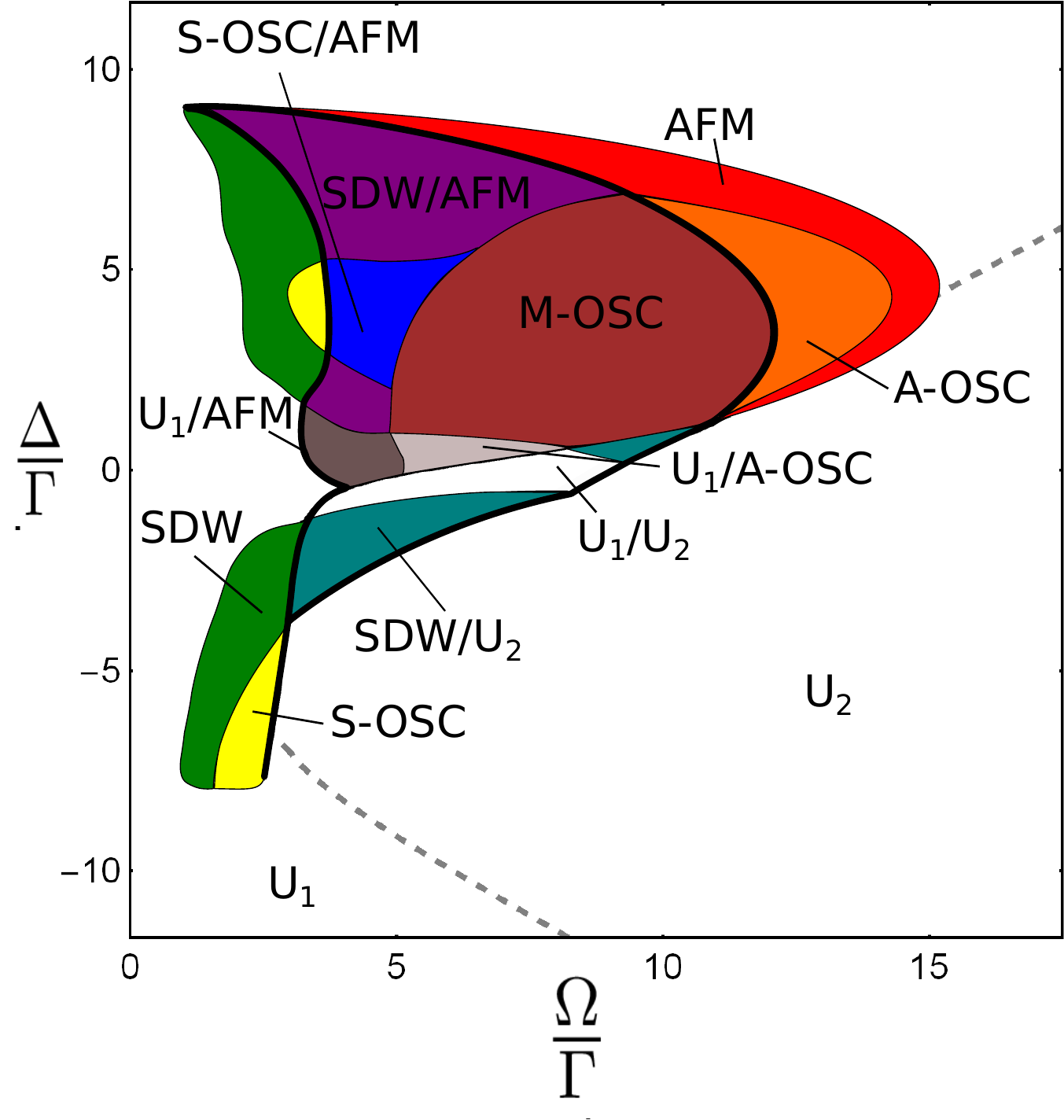}
		\vspace{-0.5cm}
		\caption{Steady state phase diagram of the system at long times. We find a variety of phases, including AFM, SDW, and OSC. Regions with two labelled phases represent bistability between those two phases. Thin lines represent second order transitions and thick lines enclose regions of bistability within which a first order transition will occur as Rabi coupling is increased. The dashed line represents an arbitrary crossover between the $U_1$ and $U_2$ phases at $S_z=-1/4$, such that we call the region with $S_z<-1/4$ the $U_1$ phase and that with $-1/4<S_z<0$ the $U_2$ phase.}
		\label{InPlanePhaseDiagram}
	\end{figure}

	By classifying the steady states of Eqs. \eqref{SpinEqs}, we can plot a phase diagram as a function of detuning and Rabi coupling. The phase diagram is shown in Figure \ref{InPlanePhaseDiagram}. To calculate the phase diagram, we find and analyse the linear stability of all the uniform and antiferromagnetic solutions of Eqs. \eqref{SpinEqs}, which determines most boundaries in the phase diagram as well as regions of bistability. To support our stability analysis, we evolve the full dynamics of Eqs. \eqref{SpinEqs} to the long-time limit (up to $t\Gamma=350$) to confirm the uniform and antiferromagnetic phases and also to determine the resultant phase when the uniform or antiferromagnetic phases become unstable. This allows us to define the boundaries between SDW and OSC phases and to check that the wavevector causing instability of the uniform solutions, $q$, has the same periodicity as the SDW phases that emerge in the full dynamics. Finally, whenever the instability wavevector is of the form $q a=2\pi/n$ where $n$ is a integer with $1<n\leq10$, we also simulate the dynamics in a sublattice ansatz, which involves reducing Eqs. \eqref{SpinEqs} to $n$ sites which repeat periodically throughout the full lattice.
	
	For the time evolution, we simulate system sizes of up to $200$ spins with periodic boundary conditions and use an initial condition of $(S_x,S_y,S_z)=(0,0,-1/2)$, which is most experimentally relevant as it represents all the atoms in the their groundstate. Throughout the text, $(S_x,S_y,S_z)=(0,0,-1/2)$ will also define our use of the term `groundstate'. We do, however, consider other initial conditions in certain regimes to check for bistability.
	
	Our analysis shows that many different long-time phases occur in the system. The simplest of these are spatially uniform phases. At low Rabi coupling, for all detuning values, the system lies close to the groundstate with $S_z\rightarrow-1/2$ as $\Omega/\Gamma \rightarrow0$, which we denote as the $U_1$ phase. At high Rabi coupling, the system lies in a state with a small value of $S_z$, where $S_z\rightarrow0$ as $\Omega/\Gamma \gg1$. We denote this as the $U_2$ phase. Both of these uniform phases are solutions of the full quantum system \cite{Kraus2008} in the limit of low/high Rabi coupling respectively.
	
	At $|\Delta/\Gamma|\gg 1$, the $U_1$ phase smoothly crosses over into the $U_2$ phase as the Rabi coupling is increased. However, when $|\Delta/\Gamma|\lesssim 1$ we find phase a sharp first order transition between the $U_1$ and $U_2$ phases, which occurs within a region of $U_1$-$U_2$ bistability.
	
	The uniform phase behaviour is analogous to a liquid-gas phase diagram where the $U_1$ phase can be considered the high density liquid phase and the $U_2$ phase as low density gaseous phase. The first order transition at $|\Delta/\Gamma|\lesssim 1$ and smooth crossover at $|\Delta/\Gamma|\gg 1$ are then similar to the liquid-gas transitions where detuning and Rabi coupling take the role of pressure and temperature respectively. In Figure \ref{InPlanePhaseDiagram}, for $|\Delta/\Gamma|\gg 1$, we define an arbitrary crossover between the $U_1$ and $U_2$ phases by the condition $S_z=-1/4$, so one can consider the $U_1$ phase to be defined as $S_z<-1/4$ and the $U_2$ phase as $-1/4<S_z<0$. When the magnitudes of detuning and Rabi coupling are comparable to the interaction between nearest neighbour sites ($|V_{12}/\Gamma|=5.32$), we also find additional non-trivial phases, which we now discuss in more detail.
	
		\begin{figure*}
			\hspace*{-0cm}
			\includegraphics[scale=0.6,clip,angle=0]{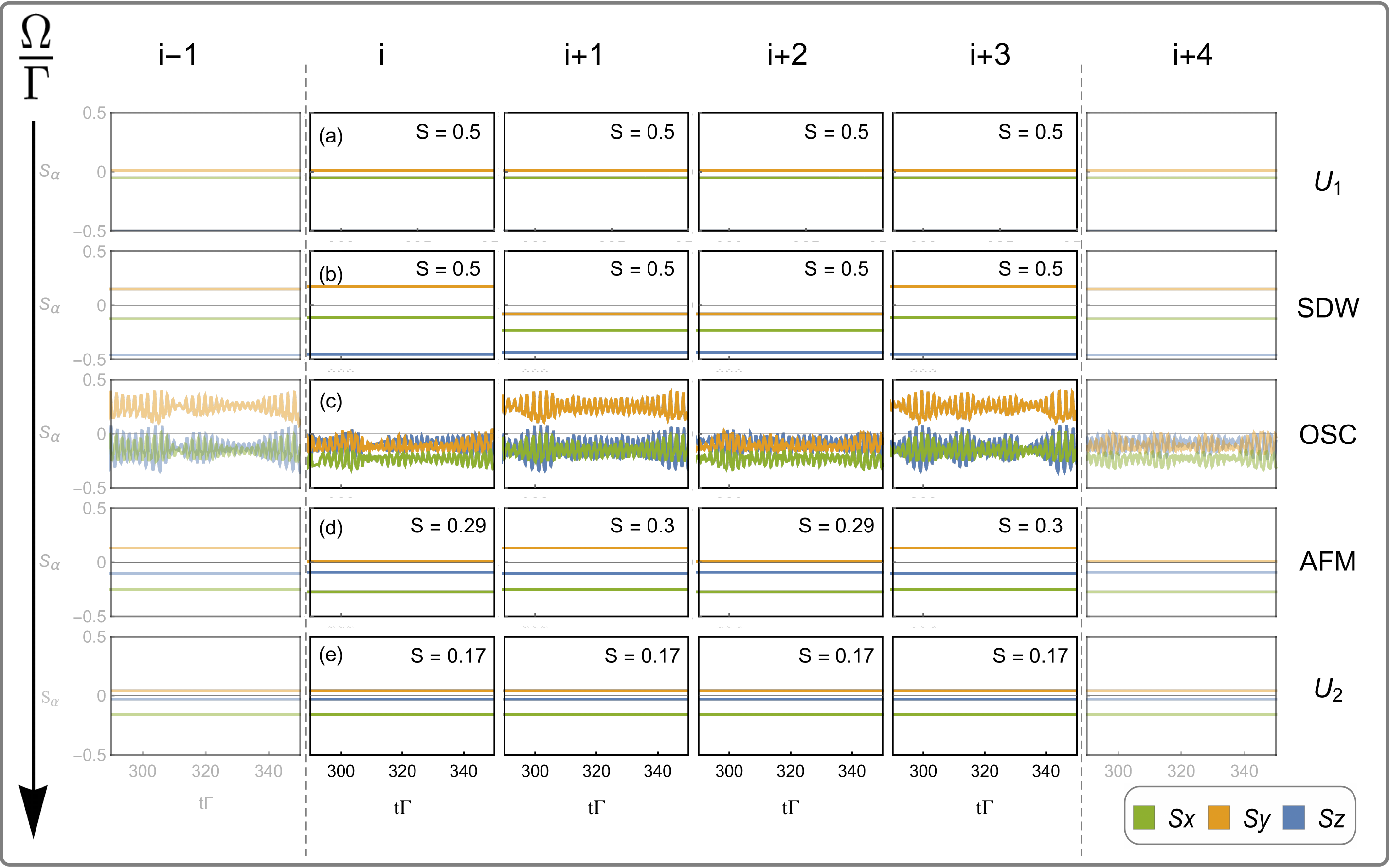}
			\vspace{-0cm}
			\caption{Examples of the spin dynamics for $\Delta/\Gamma=1.75$ for a series of sites $i$, $i+1$, $i+2$ etc. in the 1D chain. The value, $S=\sqrt{S_x^2+S_y^2+S_z^2}$, is the spin magnitude. At low Rabi coupling, there is a spatially uniform $U_1$ phase where the spins lie close to the groundstate. As the Rabi coupling increases, a SDW phase with $q a=2\pi/4$ develops, then an oscillatory phase and then an AFM phase. Finally, at high Rabi coupling we have the $U_2$ phase where the spins lie in a mixed state and the spin magnitude decreases.}
			\label{SDWvsfAFMvsLC}
		\end{figure*}
	
	\label{sec:Negative Detuning}
	\textit{Negative Detuning} - For $\Delta/\Gamma<0$, the uniform phase becomes unstable to perturbations with wavevectors in the range $0<q a <\pi$. This breaks the translational invariance of the system, and results in the formation of Spin Density Wave (SDW) phases, where the spin orientation smoothly changes across the lattice with a period set by the instability wavevector [see Figure \ref{SDWvsfAFMvsLC} (b)]. The magnitude of the instability wavevector, and hence period of the SDW, varies with detuning and Rabi coupling, moving from minimal values of around $2\pi/10$ at strong negative detuning to larger values of around $2\pi/4$ near zero detuning. 
		
	As well as the $U_1$-$U_2$ bistability mentioned earlier, we find SDW-$U_2$ bistability, where a first order transition will occur between the SDW and $U_2$ phases. Where this transition occurs and what phase the system ends up in within the bistability region depends on the initial conditions. Regimes of bistability have been found in other systems \cite{Chan2015,Wilson2016,Olmos2014,Lee2011,Lee2013} and have been observed experimentally in hot vapour gases \cite{Carr2013}. 
	
	Within certain ranges of detuning and Rabi coupling, the SDW phase can develop into an oscillatory (S-OSC) phase which persists into the long-time limit and breaks both spatial symmetry and time-translational symmetry. Oscillations, commonly referred to as limit cycles, have been reported in similar studies \cite{Chan2015,Wilson2016,Lee2011}. In contrast to the studies in \cite{Chan2015,Wilson2016,Lee2011}, we find that our limit cycles are noisy and appear chaotic, which indicates they are unstable to perturbations. We also find that while the SDW and $U_2$ can be bistable, no such bistability appears to exist between the S-OSC and $U_2$, which is possibly a consequence of the unstable nature of the oscillations. Because no such bistability exists, there is an immediate first order transition between the S-OSC and the $U_2$ phase as Rabi coupling is increased.

	\label{sec:Positive Detuning}
	\textit{Positive Detuning} - For $\Delta/\Gamma>0$, we again find the $U_1$ phase becomes unstable to perturbations, forming a SDW. However, whereas for negative detuning the $U_1$ phase only became unstable to one or two perturbations at a time, now the $U_1$ phase becomes unstable to a range of wavevectors as Rabi coupling is increased. The wavevector that causes the largest instability (indicated by the largest positive eigenvalue in the linear stability analysis) determines the period of the resultant SDW.  The wavevectors still vary with detuning and Rabi coupling, but are larger than for negative detuning, with values in the range $2\pi/3\leq q a\leq\pi$. At $q a=\pi$, the SDW becomes a canted AFM phase. We find that in certain regimes, there are additional AFM phases that can be bistable with the SDW phase.
	
	As for negative detuning, we find another oscillatory phase develops across a large range of Rabi coupling and detuning. At low Rabi coupling, noisy oscillations emerge from the SDW, forming an S-OSC phase, whilst at high Rabi coupling, the oscillations have a clear antiferromagnetic order (denoted A-OSC). At intermediate values of Rabi coupling, the oscillations take on a frustrated antiferromagnetic order due to the mixing of SDW and AFM solutions. This is also accompanied by regions of SDW-(A-OSC) bistability or small regions of (S-OSC)-(A-OSC) bistability. We do not show the boundaries between these regions but instead denote this mixture of phases as M-OSC for mixed oscillation. The boundary of the M-OSC region is defined by where the AFM phase becomes unstable or where the SDW phase disappears. 
	
	Oscillations for positive detuning with an antiferromagnetic nature have already been observed in a similar model with local dissipation \cite{Wilson2016}, including a frustrated AFM phase which seems related to our M-OSC phase. However, the S-OSC region appears to be new and also our results show a much larger region of AFM oscillation, with oscillations that contain many beat frequencies. Figure \ref{SDWvsfAFMvsLC} (c) shows an example of the AFM oscillation. 
	
\begin{figure}
		\hspace*{0 cm}
		\includegraphics[scale=0.6,clip,angle=0]{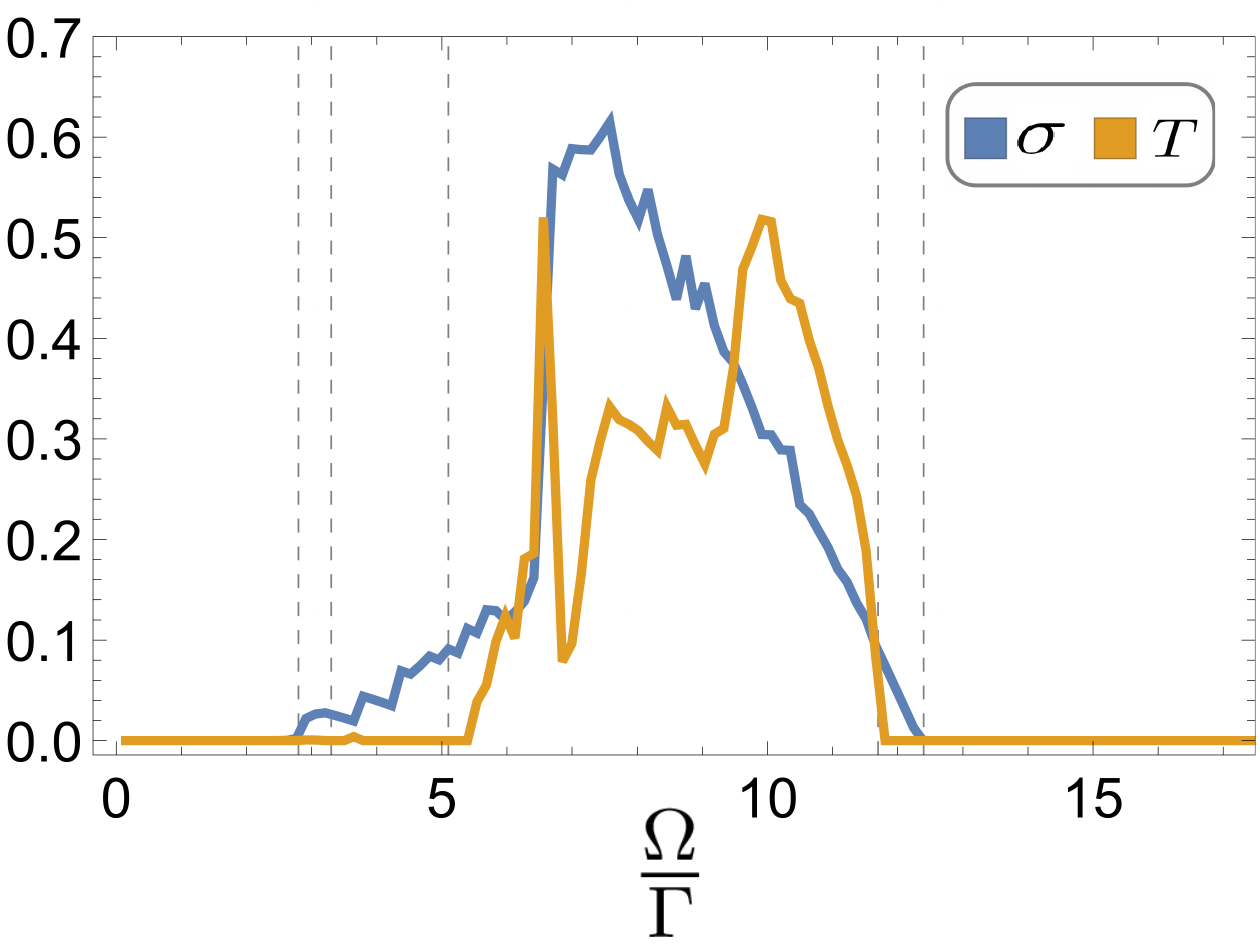}
		\vspace{-0cm}
		\caption{Evolution of the order parameters $\sigma$ and $T$, given by Eqs. \eqref{orderparam} and \eqref{orderparamT} respectively, as a function of Rabi coupling for $\Delta/\Gamma=1.75$. The SDW phase emerges via a second order transition from the $U_1$ phase at  $\Omega/\Gamma=2.8$. (Note the jagged structure at low Rabi coupling is a finite size effect, owing to competing SDW wave vectors.) The SDW then forms into an S-OSC phase and later undergoes a first order transition at $\Omega/\Gamma=6.6$ to the A-OSC phase within the M-OSC region. As the Rabi coupling is increased, the system moves to the AFM phase, where a further second order transition occurs between the AFM and $U_2$ phase at $\Omega/\Gamma=12.4$. The dashed lines indicate the crossings into the SDW, SDW/AFM, OSC, AFM and $U_2$ regions respectively.}
		\label{Order Parameter}
		\end{figure}

	In Figure \ref{Order Parameter}, we show examples of most of the phase transitions occurring within the phase diagram by simulating the full dynamics in the same parameter range as in Figure \ref{SDWvsfAFMvsLC}. To study the phase transitions, we calculate the order parameter
		\begin{equation}\label{orderparam}
		\sigma=\frac{1}{N}\sum_{i}^{N}(\bar{\textbf{S}}-\textbf{S}_i)^2,
		\end{equation}
	where $\textbf{S}_i=(S^x_i,S^y_i,S^z_i)/S$, $S=\sqrt{(S_i^x)^2+(S_i^y)^2+(S_i^z)^2}$ and $\bar{\textbf{S}}=\sum_{j}^{N} \textbf{S}_j/N$ is the average spin. We also calculate the order parameter 
	\begin{equation}\label{orderparamT}
	T=\frac{1}{N\tau}\sum_{i}^{N}\int_{0}^{\tau}(\textbf{S}_i(t_f)-\textbf{S}_i(t_f+t))^2dt,
	\end{equation}
	where $\tau=200/\Gamma$ and $t_f=700/\Gamma$, which is well into the long-time limit. The order parameter $\sigma$ takes non zero values when the phase breaks translational symmetry such as in the SDW phase and $T$ takes non zero values when the phase breaks time-translational symmetry such as in the OSC phase.
	
	We see from Figure \ref{Order Parameter} that the SDW phase emerges via a second order transition from the $U_1$ phase and then becomes unstable via another second order transition to form an S-OSC phase. The S-OSC phase then undergoes a first order transition to the A-OSC phase within the M-OSC region, leading to a sharp jump in $\sigma$ and a discontinuity in $T$. As the Rabi coupling increases, the frustration in the A-OSC phase decreases which leads to an increase in the temporal order parameter. Eventually, the OSC phase transitions to the AFM phase, which then disappears via another second order transition to the $U_2$ phase. Note that the order parameters shown here won't show a transition crossing into the M-OSC region because the M-OSC boundary is determined by when the AFM solution becomes unstable as determined by the linear stability analysis.

%owing to competing SDW wave vectors.)	After entering the M-OSC region,
	\subsection*{Explanation of Features}
	\label{sec:explanationoffeatures}
	Many aspects of the phase diagram presented here are found also for a nearest neighbour XY model with local dissipation, studied in Ref. \cite{Wilson2016}. However, there are also clear differences that arise due to the long range interactions and nonlocal dissipation. In particular, we find larger regimes of uniform phase instability at low Rabi coupling, which leads to a greater emergence of spin density waves. To understand this difference more, we study the stability of the uniform phases for systems with local and nonlocal dissipation. See Appendix A for details.
	
	\begin{figure}
		\hspace*{-0.5cm}
		\includegraphics[scale=0.33,clip,angle=0]{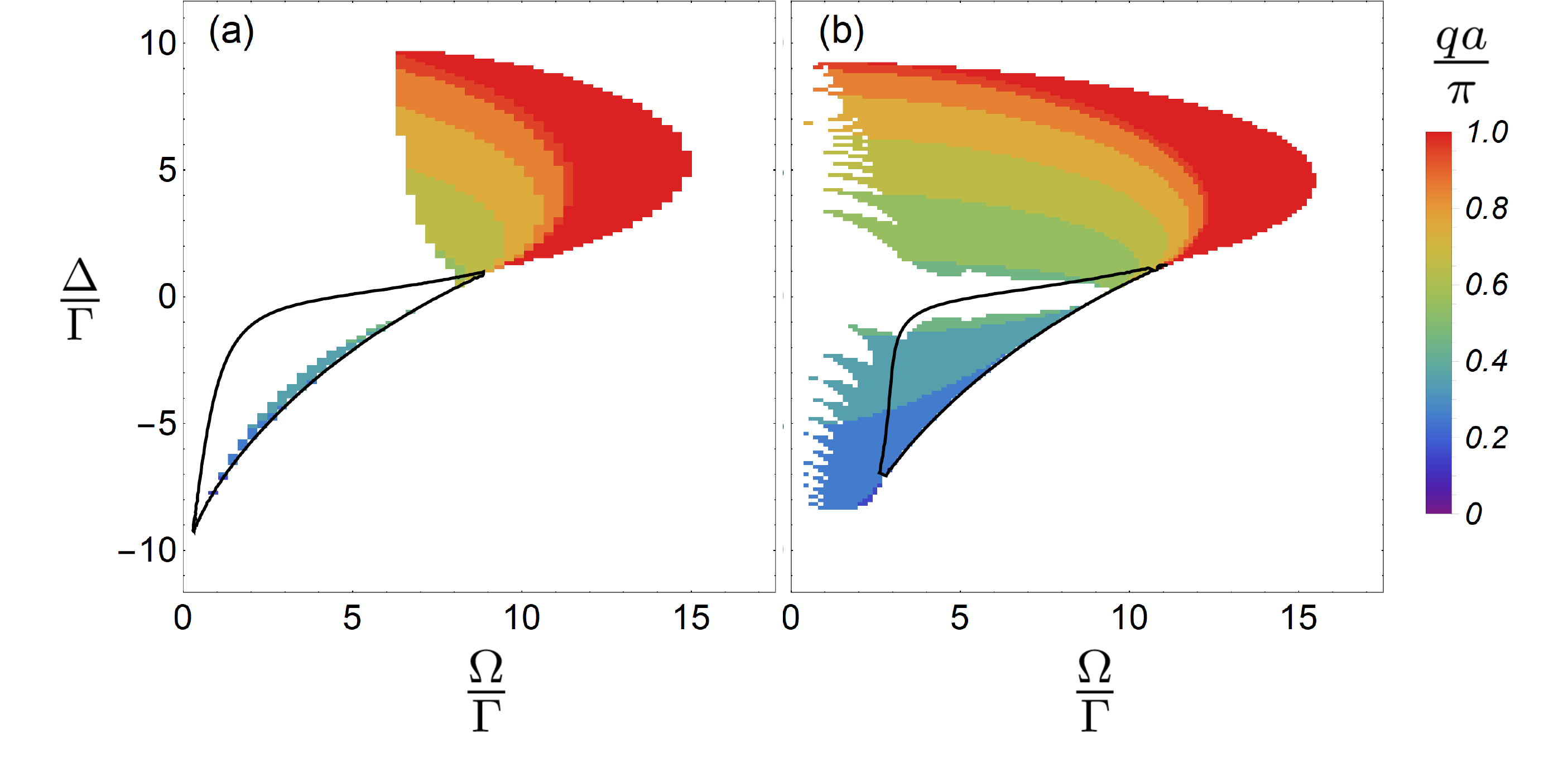}
		\vspace{-0.5cm}
		\caption{Plot of the instability wavevector of the uniform solution for systems with (a) local dissipation and (b) nonlocal dissipation. The colour represents the magnitude of the wavevector as a multiple of $\pi$ and the black line encloses the region where multiple uniform solutions exist. We can clearly see that the system with nonlocal dissipation has a larger range of instability than the system with local dissipation at lower Rabi coupling. Note that the `finned' structure at low Rabi coupling is a finite size effect.}
		\label{Kspace difference}
	\end{figure}
	
	Figure \ref{Kspace difference} shows the instability of the uniform state for a system with local and nonlocal dissipation with the black line showing the region where three uniform solutions exist. Within this region, we show the instability of only one uniform solution (the $U_1$ phase) as one solution is always stable (the $U_2$ phase) and the other unnamed phase is always unstable to perturbations with wavevector $qa=0$. We see that both systems share similarities, such as the region of multiple uniform solutions occurring at negative detuning and the same overall shape of the instability-$U_2$ crossover. The asymmetry of the phase diagram structure across the detuning range is due to the interactions, which results in a mean-field shift of the two-level transition energy. Both bistability and smallest increase in Rabi coupling needed to move from $U_1$ to $U_2$ occurs when the detuning begins to compensate for the energy shift from the interactions, bringing the drive back on resonance again. This is perhaps easiest to see from Eqs. \eqref{SpinEqs}, where in the uniform picture $dS_{x/y}/dt\sim\pm(\Delta+2S_z\sum_{i \neq 0}^{N}V_{i0})S_{y/x}$ with $\sum_{i\neq 0}^{N}V_{i0}/\Gamma=-12.4$. Considering that $-1/2\leq S_z \leq 0$, we see resonance occurs when $\Delta/\Gamma\leq0$. Specifically for the $S_z=-1/4$ contour in the phase diagram, resonance occurs at $\Delta/\Gamma=-6.2$ which is approximately where the lowest Rabi coupling is needed to cross the contour.
	
	In both systems, we find also that SDW and AFM phases can form, with a similar arrangement of instability wavevectors for positive and negative detuning. However, for the system with nonlocal dissipation, the SDW/AFM regions are larger and extend to lower Rabi coupling. To explain this, we elaborate on the linear stability analysis of the uniform phases. By linearising  Eqs. \eqref{SpinEqs} about a uniform steady state, we find the resultant matrix equation to be given by
	
	\begin{equation}\label{StabilityMatrix}\hspace*{-0.5cm}
	\begin{split}
	\frac{d}{dt} \begin{pmatrix}\delta S_z\\ \delta S_y \\\delta S_x \end{pmatrix} =\begin{pmatrix}-\Gamma & \Omega+f(q)  & g(q) \\ -\Omega+h  & -\tilde{\Gamma}(q)/2 & \tilde{\Delta}(q)\\ I  & -\tilde{\Delta}(q) & -\tilde{\Gamma}(q)/2\end{pmatrix}\begin{pmatrix}\delta S_z\\ \delta S_y \\\delta S_x\end{pmatrix}.
	\end{split}
	\end{equation}
	where $f(q)=-2S_y(\epsilon_{\Gamma}(0)+\epsilon_{\Gamma}(q))+2S_x(\epsilon _V(q)-\epsilon _V(0))$,
	$h=2S_y\epsilon_{\Gamma}(0)+2S_x\epsilon _V(0)$, $g(q)=-2S_x(\epsilon_{\Gamma}(q)+\epsilon_{\Gamma}(0)) +2S_y(\epsilon_V(0)-\epsilon_V(q))$ and $I= -2S_y\epsilon_V(0) +2S_x\epsilon_{\Gamma}(0) $. The functions $\epsilon_V(q)=\sum_{l\neq0}^{N}V_{l0}\exp(iqr_{l0})$ and $\epsilon_{\Gamma}(q)=\sum_{l\neq0}^{N}\Gamma_{l0}\exp(iqr_{l0})$ are the dispersion relations, with $\epsilon_V(0)/\Gamma=-12.4$ and $\epsilon _{\Gamma}(0)/\Gamma=2.9$ for our system, and $q$ being the momentum fluctuation. Note that the $q$ used here is the same $q$ used to classify the SDW phases earlier in the text.
	
	From Eq. \eqref{StabilityMatrix}, we see that the interactions modify the detuning, resulting in $\tilde{\Delta}(q)=\Delta+2S_z\epsilon_V(q)$. We also see that whilst the nonlocal dissipation alters the off-diagonal elements of the matrix, the crucial difference is the fluctuations in $S_{x/y}$ have an effective damping, $\tilde{\Gamma}(q)/2=\Gamma/2-S_z\epsilon_{\Gamma}(q)$. This effective damping is a direct consequence of the nonlocal dissipation and cannot occur in a system with local dissipation, where the $S_{x/y}$ fluctuations would always experience a fixed decay rate of $\Gamma/2$. 
	
\begin{figure}
		\hspace*{-0.5 cm}
		\includegraphics[scale=0.38,clip,angle=0]{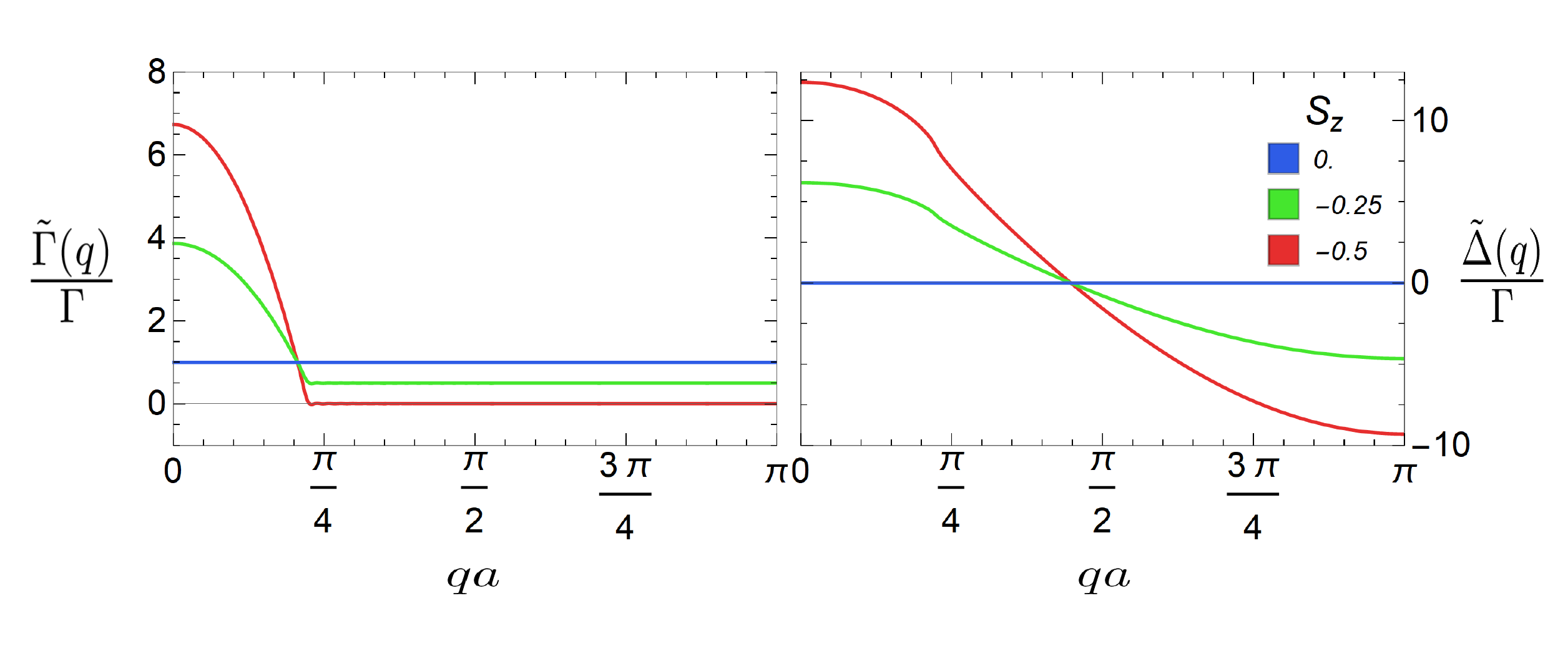}
		\vspace{-0.5cm}
		\caption{Effective damping and detuning of the $S_{x/y}$ fluctuations for different values of $S_z$. While modification of the detuning by interactions occurs for any system with interactions, the modification of the decay rate is a consequence of nonlocal dissipation. We see that the effective damping quickly becomes subradiant for higher values of $q a$, but eventually becomes equal to the onsite decay as $S_{z}$ increases.}
		\label{Gamma and delta eff}
	\end{figure}

	In Figure \ref{Gamma and delta eff}, we plot $\tilde{\Gamma}(q)$ and $\tilde{\Delta}(q)$ as functions of wavevector, $q$, for different values of $S_z$. Focussing on $\tilde{\Gamma}(q)$, we see that it quickly drops to minimal values for $qa \geq 2\pi/10$, which means the $S_{x/y}$ fluctuations experience a subradiant decay rate. By having a reduced dissipation, fluctuations can grow at much lower Rabi coupling compared to the local dissipation model, resulting in instabilities and the formation of SDW phases. The fact that high momentum fluctuations are subradiant can be understood as a result of destructive interference between the dipoles, which begin to oscillate out of phase, inhibiting photon emission and therefore trapping excitations in the system. As the Rabi coupling is increased, $S_z$ will decrease in value and $\tilde{\Gamma}(q)$ eventually tends to $\Gamma$. This results in the similarity between the local and nonlocal dissipation instability plots at higher Rabi coupling, as the effects of nonlocal dissipation become negligible. 

	Examining the characteristic polynomial from the matrix in Eqs. \eqref{StabilityMatrix} gives more insight into the phase diagram structure.
	We find that stability of the uniform phase is determined by the sign of the expression
	
	\begin{equation}\label{A0}
	\begin{split}
	A_0=&\frac{\tilde{\Gamma}}{2}\left[\frac{\tilde{\Gamma}(q)}{2}\Gamma+\Omega(\Omega-h+f(q))-Ig(q)-f(q)h\right]+\\
	&\tilde{\Delta}(q)\left[\tilde{\Delta}(q)\Gamma-\Omega(I+g(q))+g(q)h-If(q)\right].
	\end{split}
	\end{equation}
	Full details on why this is the case is given in Appendix B. Here we see that the expression in the first set of brackets is multiplied by $\tilde{\Gamma}(q)$. If $\tilde{\Gamma}(q)\approx0$, this means the sign of $A_0$, and hence the stability of the uniform solution, is determined by the sign of $\tilde{\Delta}(q)$ and the expression $\tilde{\Delta}(q)\Gamma-\Omega(I+g(q))+g(q)h-If(q)$. By looking at the dispersion $\epsilon_V(q)$ plotted in Figure \ref{Gamma and delta eff}, we can see that the sign of $\tilde{\Delta}(q)$ depends on the value of the detuning and the momentum wavevector. If the detuning is positive, then only wavevectors between $2\pi/4<q a<\pi$ can cause $\tilde{\Delta}(q)<0$ and hence instabilities, whereas if the detuning is negative, then only wavevectors with $0<q a<2\pi/4$ can cause instabilities. This therefore explains the ordering of spin density waves for negative and positive detuning.
	
	\label{Oscillation Explanation}
	We also found the emergence of two OSC phases in our phase diagram, one for $\Delta/\Gamma>0$ and one for $\Delta/\Gamma<0$. As mentioned earlier, aspects of the OSC phase for $\Delta/\Gamma>0$ have been seen in the system with local dissipation, whereas the OSC phase for $\Delta/\Gamma<0$, which occurs on the SDW-$U_2$ boundary, is new and a consequence of nonlocal dissipation. We would intuitively expect oscillations to occur on the boundaries between two phases with different spin orientations where the orientation of the spins is susceptible to change direction \cite{Chan2015} and so can be easily driven. Therefore, this new OSC phase is linked to the emergence of the SDW phases at negative detuning.
	
	The oscillations within this phase appear to be noisy and chaotic. We study the emergence and dynamics of the oscillations in more detail by employing a sublattice ansatz. Analysing the stability of the sublattice solution, we determine that the oscillations arise from Hopf bifurcations \cite{Liu1994} in the SDW phase, which lead to stable limit cycles. Checking the stability of these limit cycles using classical Floquet analysis, we find that they become unstable to perturbations with wavevectors not allowed in the sublattice system. Whilst the underlying cause of this is unclear in detail, one can imagine that if one were to drive and populate several highly subradiant modes, then the system would behave as a closed driven XY model with dipole couplings which has been shown to have unstable noisy oscillations \cite{Parmee2017}.
	
	We now focus on the OSC phases for $\Delta/\Gamma>0$, again employing a sublattice ansatz and Floquet analysis. We find at low Rabi coupling, the SDW phase can become unstable, giving rise to oscillations that are mostly noisy and chaotic, just as for the OSC phase at $\Delta/\Gamma<0$. However, there are additional AFM phases for positive detuning that can be bistable with the SDW and S-OSC phases and these AFM solutions also become unstable as the Rabi coupling is increased. This gives rise to the M-OSC phase where both forms of oscillation can mix or  the SDW and A-OSC phases mix. There are also regimes of SDW-(A-OSC) bistability and (S-OSC)-(A-OSC) bistability. As the wavevector of the SDW tends to $\pi$ with increasing Rabi coupling, eventually only the A-OSC phase exists. Therefore we find that nonlocal dissipation gives rise to two components for the OSC phase; firstly, the emergence of an S-OSC phase which does not occur in the system with local dissipation and secondly a region of A-OSC phase which is much larger than in the system with local dissipation.

	\section{Beyond Mean Field}
	\label{sec:beyondmeanfield}
	Throughout our analysis, we have employed a mean-field approximation. At very low Rabi coupling, this approximation captures the full quantum model because the system lies close to the ground state, with $\sum_{i}^{N}S_i^z\approx-N/2$ being nearly conserved due to the Hamiltonian. Single excitations can then be viewed classically due to the large effective spin, allowing the system to be mapped to coupled oscillators. However, for higher Rabi coupling where the interesting phases emerge, quantum effects will be more significant. Mean-field theory is expected to be valid for a higher effective co-ordination number where quantum fluctuations can cancel on average. Therefore our results in 1D are most susceptible to quantum fluctuations, although long-range interactions help increase the effective co-ordination number. Despite this, mean-field theory is still expected to capture some aspects of the full quantum system.
	
\begin{figure}
		\hspace*{0cm}
		\includegraphics[scale=0.6,clip,angle=0]{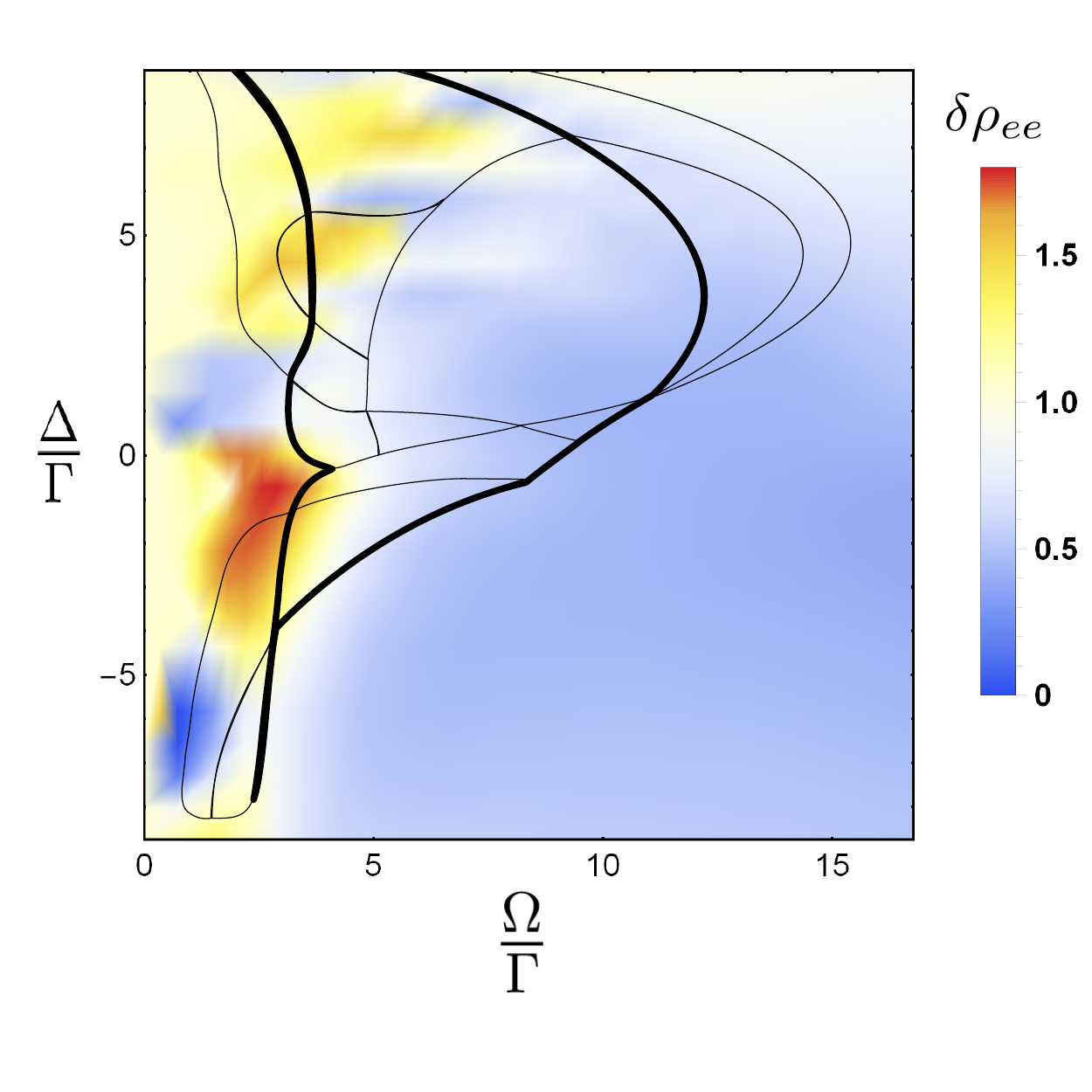}
		\vspace{-0.5cm}
		\caption{The normalised excitation fluctuations for $N=10$ spins with periodic boundary conditions. The fluctuations peak at the onset of bistability.}
		\label{FluctuationPlot}
	\end{figure}

	In regimes where mean-field theory predicts bistability, we expect a unique steady state in the full quantum regime \cite{Schirmer2010} and a smooth crossover between the $U_1$ and $U_2$ phases \cite{Clark2016} rather than a sharp transition. However, signatures of bistability can be found in the excitation density fluctuations. On the approach to the steady state, the excitation density, $\rho_{ee}=\langle S^z\rangle+1/2$, has been shown to fluctuate between the two bistable states using Quantum Monte Carlo Wavefunction methods \cite{Olmos2014,Wilson2016,Lee2012a,Ates2012a}. This fluctuating leads to bimodality in the excitation distribution and a peak in the normalised fluctuations, $\delta\rho_{ee}=\sum_{i, j}^{N}\left(\langle S^z_iS^z_j\rangle-\langle S^z_i\rangle\langle S^z_j\rangle\right)/\rho_{ee}$. To calculate $\delta\rho_{ee}$, we first find the steady state by writing  Eq.\eqref{MasterEq} in matrix form such that $d\rho/dt=\mathcal{L}\rho$ and then find the eigenstate of the Liouvillian matrix $\mathcal{L}$ with an eigenvalue of zero. The corresponding eigenvector is the steady state density matrix \cite{Navarrete-Benlloch} and from that we can calculate the excitation fluctuation. Figure \ref{FluctuationPlot} shows a plot of $\delta\rho_{ee}$ for a system of $N=10$ spins with periodic boundary conditions. We find a peak in $\delta\rho_{ee}$ near the onset of both bistability regions. Similar results have been seen in \cite{Olmos2014} for the uniform bistability region. 	
	
\begin{figure}
		\hspace*{-0.7cm}
		\includegraphics[scale=0.5,clip,angle=0]{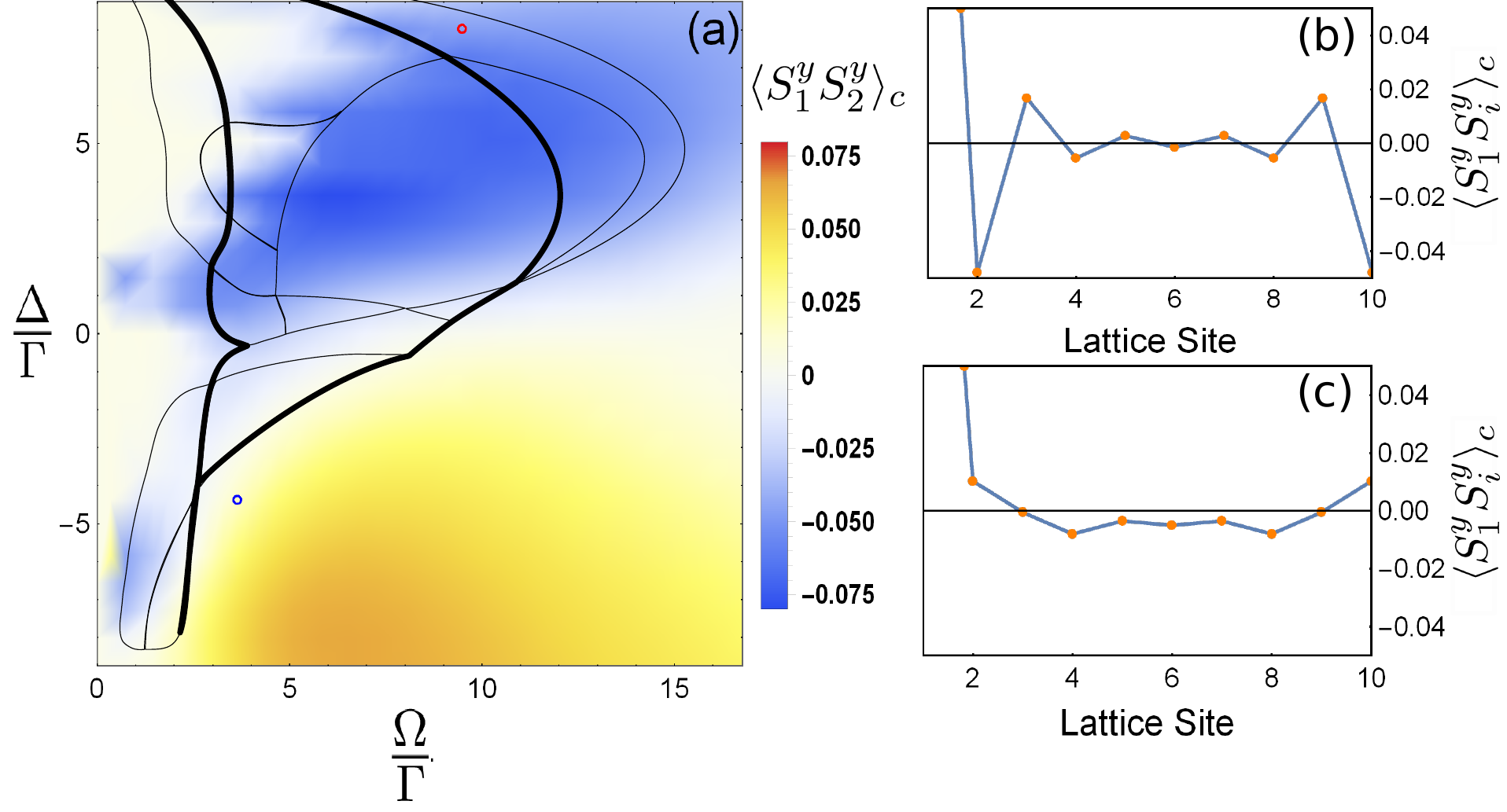}
		\vspace{-0.5cm}
		\caption{(a) Connected correlator $\langle S_1^yS_2^y \rangle_c$ as a function of Rabi coupling and detuning for $N=10$ spins on a chain with periodic boundary conditions. We see the correlation is negative for $\Delta/\Gamma>0$ and positive for $\Delta/\Gamma<0$. The black lines show the mean-field phase diagram boundaries. The insets (b) and (c) show examples of $\langle S_1^yS_i^y \rangle_c$ along the spin chain at the points indicated by the red and blue circles in (a) respectively.}
		\label{correlation plot}
	\end{figure}

	Steady state correlation functions should retain order corresponding to the wavevector of instability, although losing long range order \cite{Lee2011, Wilson2016,Joshi2013}. Therefore we also calculate the connected correlator, $\langle S_i^yS_j^y \rangle_c\equiv\langle S_i^yS_j^y \rangle-\langle S_i^y\rangle\langle S_j^y\rangle $, for $N=10$ spins on an chain with periodic boundary conditions. In Figure \ref{correlation plot}, we plot $\langle S_1^yS_2^y \rangle_c$ across the entire phase diagram, although the same results hold for any spin in the chain due to translational symmetry.
	Our results show that the correlations do indeed lose long range order, but take an antiferromagnetic nature for $\Delta/\Gamma>0$. For $\Delta/\Gamma<0$, spins become more positively correlated with their nearest neighbours in the region where the uniform phase persists, which agrees with the mean-field phase diagram. Therefore our quantum checks indicate that aspects of the mean-field theory should persist in smaller quantum systems.
	
	\section{Discussion}
	\label{sec:discussion}
\begin{figure}
		\center
		\includegraphics[scale=0.7,clip,angle=0]{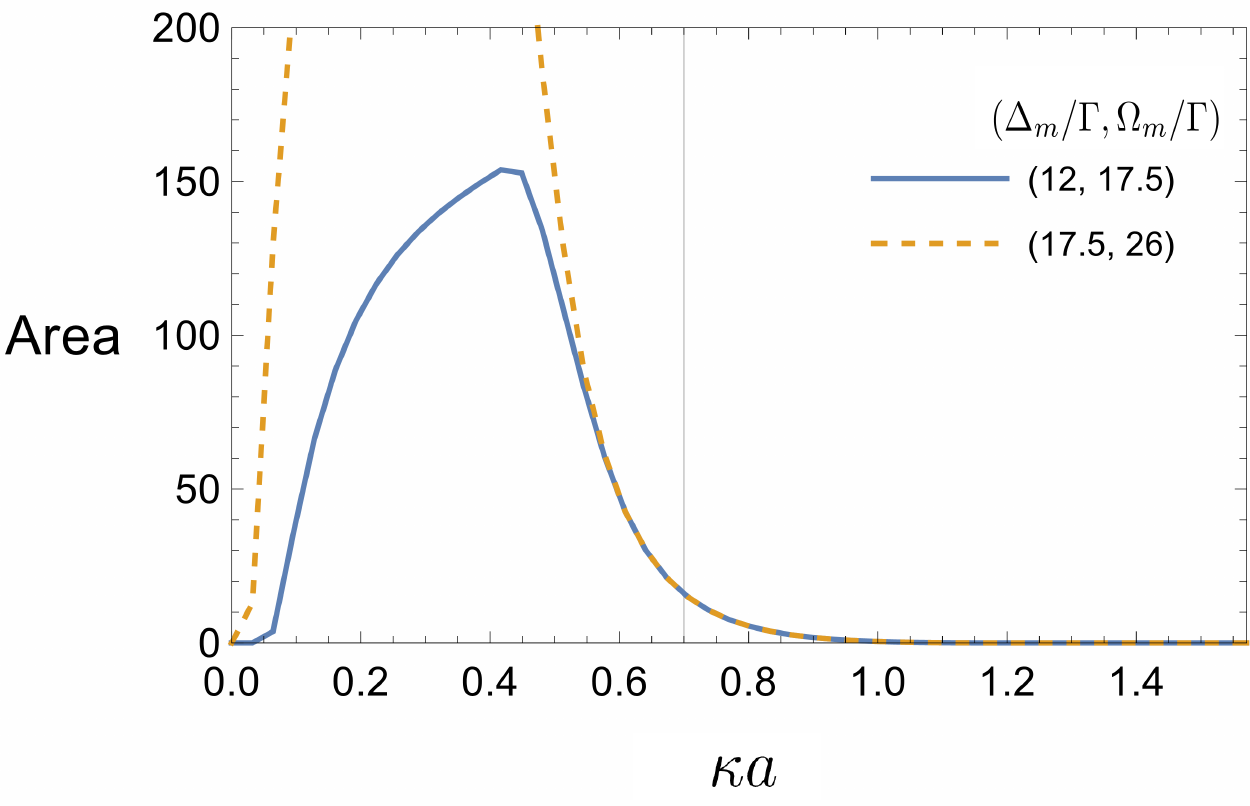}
		\vspace{-0cm}
		\caption{Area of the multiple uniform solution region (shown enclosed by the black line in Fig \ref{Kspace difference}) as a function of lattice spacing. For $\kappa a\geq1.2$, the area is nearly zero and we expect only uniform phases to exist beyond this limit. It should be noted that at low lattice spacing, the interaction strength diverges, causing the area of multiple uniform solutions to artificially peak and then decrease to zero within any fixed range of $\Delta/\Gamma$ and $\Omega/\Gamma$. We emphasise this by showing two curves for the ranges $-\Delta_m/\Gamma\leq\Delta/\Gamma\leq\Delta_m/\Gamma$ and $0\leq\Omega/\Gamma\leq\Omega_m/\Gamma$. At higher $\kappa a$ the curves overlap so the cutoff doesn't affect our results. The grey line shows the choice of lattice spacing of $\kappa a=0.7$ used in this study.}
		\label{discplot}
	\end{figure}

	We have explored the phase diagram of a ensemble of two-level systems under an external drive and with resonant dipole-dipole interactions, finding the emergence of SDW, AFM, OSC and bistable phases and determining how the formation of these phases relates to nonlocal dissipation. To realise such a system experimentally, Sr atoms can be used, with the two level transition between the $^3P_0$ and $^3D_1 (m=0)$ levels \cite{Olmos2013}. This transition has a transition wavelength of $\lambda=2.6\mu $m and would require a lattice spacing of $a=289.6$nm to achieve $\kappa a=0.7$. Other lattice spacings and atomic species may be used, as we expect many of our results to extend to nearby values of $\kappa a$. We do find however, that beyond a certain lattice spacing, the interaction between spins becomes insignificant. A good indicator of where this cut off occurs can be determined by looking at the region of multiple uniform solutions, enclosed by the black line in Fig. \ref{Kspace difference}. In Figure \ref{discplot}, we plot how the area of this region changes as a function of $\kappa a$. We find that as $\kappa a$ increases, the area decreases and eventually disappears at $\kappa a\approx 1.2$. Beyond this limit, we expect only uniform phases to exist. 

	In our simulations, we evolved the system to times of $t\Gamma=350$ or greater to reach the steady state. It may be the case that some phases we find are metastable with a very long decay time. Furthermore, the majority of our simulations were carried out for an initial condition of all the spins in the groundstate, though we did use other initial conditions to examine regions of bistability. We believe our analysis accounts for the majority of phases that exist in the system, but there may be other bistable/multistable phases not captured in our phase diagram that can occur for other initial conditions. Finally, we have only considered 1D systems under uniform driving. It would be interesting to see what features change in higher dimensions, different geometries and under non-uniform driving given the presence of nonlocal dissipation. While our quantum results indicated some aspects of the mean-field theory should be observable, it would also be of interest to study the full quantum system in more detail and quantify where the mean-field theory approximation may fail.
	
	\section{Conclusions}
	\label{sec:conclusions}
	We have studied the mean-field nonlinear dynamics of a 1D chain of two-level systems coupled with dipole-dipole interactions and nonlocal dissipation being driven by an external field. We determined the phases that form in the long-time limit such as antiferromagnetism, spin density waves, oscillations and phase bistabilities. We find that the nonlocal dissipation plays a key role in the emergence of these phases by coupling fluctuations in the system to different decay modes and causing a greater formation of spin density and oscillatory phases. We also find that some of the mean-field features persist in the full quantum regime.

	\section{Acknowledgements}
	This work was supported by EPSRC Grant No. EP/K030094/1. and No. EP/P009565/1. Statement of compliance with EPSRC policy framework on research data: All data accompanying this publication are directly available within the publication.
	
	\begin{appendices}
		\section*{A - Uniform Solutions}
		To determine the uniform phases in the systems with local and nonlocal dissipation, we solve the equations of motion, Eqs. \eqref{SpinEqs}, for a single site, which allows us to obtain the following cubic polynomial
		
		\begin{equation}\label{SScubic}
		\begin{split}
		&4(\epsilon_{\Gamma}(0)^2+\epsilon_V(0)^2+2\Delta \epsilon_V(0)-\Gamma \epsilon_{\Gamma}(0))S_z^2+\\
		&2(\Delta^2+\Gamma^2/4+\Omega^2/2+2\Delta \epsilon_V(0) -\Gamma \epsilon_{\Gamma}(0))S_z+\\
		&8(\epsilon_{\Gamma}(0)^2+\epsilon_V(0)^2)S_z^3+(\Delta^2+\Gamma^2/4)=0.
		\end{split}
		\end{equation}
		
		The discriminant of a cubic given by $ax^3+bx^2+cx+d$ is $b^2c^2-4ac^3-4b^3d-27a^2d^2+18abcd$. By substituting in for a, b, c and d, we can determine the number of real roots, and hence steady state solutions, of Eq. \eqref{SScubic}. If the discriminant is greater than zero, there are three solutions, while if it is less than zero, there is only one solution. Using this, we can easily find the region of multiple uniform solutions and how the area of this region changes as a function of lattice spacing as plotted in Figure \ref{discplot}. To look at the solutions for local dissipation only, we set $\epsilon_{\Gamma}(0)=0$ Eq. \eqref{SScubic}.
		
		Once we have obtained the uniform solutions, we check their stability to linear perturbations by linearising Eqs. \eqref{SpinEqs} about the uniform steady state, which gives us matrix equation Eq. \eqref{StabilityMatrix} in the main text. Once again, to look at local dissipation only, we set $\epsilon_{\Gamma}(0)=0$ in Eq. \eqref{StabilityMatrix}.

		\section*{B - Uniform Phase Stability}	
		Stability of the uniform solution comes from the eigenvalues of the characteristic polynomial, given by $A_3 \lambda^3+A_2 \lambda^2+A_1 \lambda^1+A_0$. We can formulate the Routh array of the characteristic polynomial and determine stability from the coefficients without explicit knowledge of the uniform solution \cite{Sturm}. The Routh array is given by
		\begin{table}[H]
				%\caption{Routh Array Coefficients}
				\centering % used for centering table
			\begin{tabular}{c c} % centered columns (4 columns)
				%heading
				\hline % inserts single horizontal line
				$A_3$ & $A_1$  \\ % inserting body of the table
				$A_2$ & $A_0$  \\
				$\frac{A_2A_1-A_0A_3}{A_2}$ & $0$ \\
				$A_0$ & $0$ \\
				[1ex] % [1ex] adds vertical space
				\hline %inserts single line
			\end{tabular}
			\label{table:routh} % is used to refer this table in the text
		\end{table}	
	For stability, $A_2$, $A_1A_2-A_0A_3$ and $A_0$ all need to be greater than zero for the solution to be stable. If $A_1A_2-A_0A_3$ changes sign, then the system undergoes a Hopf bifurcation as we have a row of zeroes with no sign change on either side of the row. We find numerically this does not happen for the uniform solution but does occur within the SDW/AFM phases. From the matrix equation, Eq. \eqref{StabilityMatrix}, we can calculate the values $A_0$ to $A_3$ which are given by 
	
	\begin{equation}
	\begin{split}
	A_3=&1\\
	A_2=&\Gamma+\tilde{\Gamma}(q)\\
	A_1=&\frac{\tilde{\Gamma}(q)}{2}\left(\frac{\tilde{\Gamma}(q)}{2}+2\Gamma\right)+\Omega(\Omega-h+f(q))\\
	&+\Delta^2-Ig(q)-f(q)h \\		A_0=&\frac{\tilde{\Gamma}}{2}\left[\frac{\tilde{\Gamma}(q)}{2}\Gamma+\Omega(\Omega-h+f(q))-Ig(q)-f(q)h\right]+\\
	&\tilde{\Delta}(q)\left[\tilde{\Delta}(q)\Gamma-\Omega(I+g(q))+g(q)h-If(q)\right].
	\end{split}
	\end{equation}
	
	We find that $A_2$ and $A_1A_2-A_0$ are always greater than zero, so stability is given by $A_0$ only, which explains why we analyse the expression in Eq. \eqref{A0} in the main text.
		
	\end{appendices}
	
	\nocite{*}
	%\bibliographystyle{apsrev4-1}
	%\bibliography{Paper2}
	
	%merlin.mbs apsrev4-1.bst 2010-07-25 4.21a (PWD, AO, DPC) hacked
	%Control: key (0)
	%Control: author (72) initials jnrlst
	%Control: editor formatted (1) identically to author
	%Control: production of article title (-1) disabled
	%Control: page (0) single
	%Control: year (1) truncated
	%Control: production of eprint (0) enabled
	%

\end{document}